\documentclass[seceq]{ptptex}

\usepackage{graphicx}
\usepackage{wrapft}

\newcommand{\ga}{\stackrel{>}{ _{\sim}}}

\newcommand{\ve}[1]{\mbox{\boldmath $#1$}}

\title{Numerical Relativity at the Frontier}

\author{Stuart L.\ \textsc{Shapiro}}

\inst{Departments of Physics and Astronomy, University of Illinois
 at Urbana-Champaign \\
Loomis Laboratory, 1110 West Green Street, Urbana, IL 61801}

\abst{
Numerical relativity is an essential tool for solving Einstein's
equations of general relativity for dynamical systems characterized
by high velocities and strong gravitational fields. The implementation 
of new algorithms that can solve these nonlinear equations in $3+1$ 
dimensions has enabled us to tackle many long-standing problems of 
astrophysical interest, leading to an explosion of
important new results.  Numerical relativity has been used to
simulate the evolution of a diverse array of physical systems, 
including coalescing black hole and neutron star binaries, rotating and
collapsing compact objects (stars, collisionless clusters, 
and scalar fields), and magnetic and viscous stars, to name a few. Numerical
relativity has been
exploited to address fundamental points of principle, including critical 
phenomena and cosmic censorship. It holds great promise as a 
guide for interpreting observations of 
gravitational waves and gamma-ray bursts and identifying the sources
of such radiation. Highlights of a  
few recent developments in numerical relativity are sketched in this 
brief overview.
}

\begin{document}

\maketitle

\section{Introduction}

General relativity---Einstein's theory of
relativistic gravitation---is the cornerstone of
modern cosmology, the physics of neutron stars and
black holes, the generation of gravitational
radiation, and countless other cosmic phenomena in
which strong-field gravitation is believed to play a
dominant role.  However solutions to Einstein's equations,
except for a few idealized cases characterized by
high degrees of symmetry, have not been obtained as
yet for many of the important dynamical scenarios
thought to occur in nature. Only now, with the
advent of supercomputers, is it possible to tackle
these highly nonlinear equations numerically and
to explore these scenarios in detail. That is the main
goal of numerical relativity, the art and science of
developing computer algorithms to solve Einstein's
equations for physically realistic, high-velocity,
strong-field systems. Numerical relativity also has
a pressing goal---to calculate gravitational
waveforms from promising astrophysical sources, in
order to provide theoretical templates both for the
new ground-based gravitational-wave laser interferometers
like LIGO in the US, VIRGO in Italy,
GEO in Germany, and TAMA in Japan, as well as for  space-based
interferometers such as LISA, now being planned in the US, Europe
and Japan.
                                                                                
This paper is a brief summary of a few 
recent developments in numerical relativity. It is 
far from complete and undoubtedly subject to my personal tastes 
and preferences. 
For example, my own short list of the `hottest' topics in the field 
currently consists of the following:\\
\noindent $\bullet$ Binary black holes (BBHs) \\
$\bullet$ Binary neutron stars (BNSs) \\
$\bullet$ Binary black hole-neutron stars (BBHNSs) \\
$\bullet$ Rotating relativistic stars \\
$\bullet$  Collisionless clusters \\
$\bullet$ Scalar fields \\
$\bullet$ Critical phenomena \\
$\bullet$ Cosmic censorship \\
$\bullet$ General relativistic magnetohydrodynamics (GRMHD) \\

The inspiral and coalescence of BBHs, BNSs and BBHNSs top the list, 
as they represent the 
most likely sources of gravitational waves for detection by the laser 
interferometers [See Ref.~\citen{rf:bs} for a recent review and references 
and Refs.~\citen{rf:stu05,rf:bhns,rf:pre} for 
important updates].  BNSs and BBHNSs are also strong candidates for
short-period gamma-ray burst sources. Numerical relativity is crucial for
constructing astrophysically realistic 
initial configurations of relativistic binaries in close, 
quasiequilibrium, nearly circular orbits, as well as for tracking 
their subsequent inspiral and merger. In the
case of fluid star companions, the equations of relativistic hydrodynamics 
must be integrated together with Einstein's gravitational field equations.
Numerical relativity is also important in the study 
of rotating, relativistic stars, especially NSs. 
Gravitational collapse, the dynamical evolution of stars 
subject to nonaxisymmetric instabilities (like the dynamical bar-mode 
instability\cite{rf:sbs,rf:ssbs}) and the secular evolution of stars driven by
viscosity,\cite{rf:dlss} are all being explored by numerical simulations.
Collisionless matter obeying the relativistic Vlasov equation 
has been treated in depth using the tools of 
numerical relativity.\cite{rf:st,rf:rst,rf:oc}. For example, binary clusters of
collisionless particles 
that undergo collapse prior to merger have been used to simulate 
head-on collisions of binary black holes\cite{rf:st92} and the results
have been useful in probing the geometry of coalescing event 
horizons\cite{rf:matz95}.
Scalar fields have been
employed as matter sources to study wave propagation, collapse,\cite{rf:ss} 
critical phenomena\cite{rf:cho,rf:gun} and binary BHs.\cite{rf:pre}
In the course of all of these numerical simulations, 
questions of fundamental
physics have been addressed, including critical phenomena and cosmic
censorship.\cite{rf:st91} One of the latest thrusts, 
general relativistic magnetohydrodynamics (GRMHD) in dynamical
spacetimes, promises to probe some of the astrophysically 
most important, unsolved aspects of  
magnetized stars and disks in strong gravitational 
fields.\cite{rf:bs03a,rf:dlss05,rf:ss05,rf:a05}

In the sections below I will sketch results of some recent work 
on several of the topics described above.
My emphasis will be on astrophysical systems that
involve strong gravitational fields coupled to relativistic matter sources.
My goal is not to record final, definitive solutions but rather to 
convey the flavor and breadth of current activity and the 
extent to which the latest algorithms and computational machinery 
now make it possible to obtain definitive solutions.

\section{3+1 Formalism}

\begin{figure}
       \centerline{\includegraphics[height=3.3 cm]
                                   {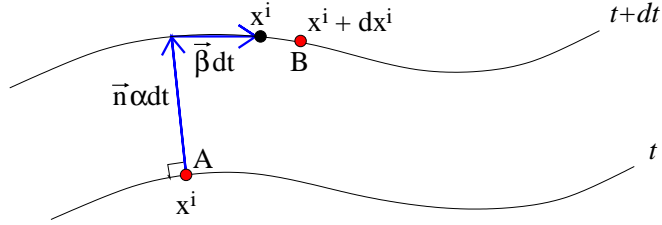}}
   \caption{3+1 decomposition of spacetime.}
   \label{fig1}
   \end{figure}

Most current work in numerical relativity is performed within the framework of
the $3+1$ decomposition of Einstein's field equations using some
adaptation of the standard ADM equations.\cite{rf:adm}. 
In this framework spacetime is
sliced up into a sequence of spacelike hypersurfaces of constant
time $t$, appropriate for solving an initial-value problem.
Consider two such time slices separated by an infinitesimal interval $dt$ as 
shown in Fig~\ref{fig1}.  
The spacetime metric gives the invariant interval between neighboring 
points $A$ and $B$ on the two slices according to
\begin{equation}
ds^2 = -\alpha^2 dt^2 +
\gamma_{ij}(dx^i+\beta^i dt) (dx^j+\beta^j dt)\ ,
\end{equation}
where $\gamma_{ij}$ is the 3-metric 
on a time slice, $\alpha$ is the lapse function determining
the proper time between the slices as measured by a time-like normal observer 
$n^a$ at rest in the slice, and $\beta^i$ is the  
shift, a spatial 3-vector 
that describes the relabeling of the spatial coordinates of points in the slice.
The gravitational field satisfies the Hamiltonian and
 momentum {\it constraint equations} 
on each time slice, including the initial slice at $t=0$:
\begin{eqnarray}
R + K^2 - K_{ij}K^{ij} &=& 16 \pi \rho
~~({\rm Hamiltonian})\ , \label{ham} \\
D_j(K^{ij} - \gamma^{ij}K) &=& 8 \pi S^i
~~({\rm momentum})\ \label{mom}.
\end{eqnarray}
Here $R=R^i{}_i$ is the scalar curvature on the slice,
$R_{ij}$ is the 3-Ricci tensor, $K_{ij}$ is the 
extrinsic curvature, $K$ is its trace, and $D_j$ is the covariant
derivative operator on the slice. The quantities $\rho$ and $S^i$ are
the mass and momentum densities of the matter, respectively; such matter source terms
are formed by taking suitable projections of the matter stress-energy tensor
$T^{ab}$ with respect to the normal observer.

A gravitational field  satisfying the constraint equations on the initial
slice can be determined at future times by integrating the 
{\it evolution equations},
\begin{eqnarray}
\partial_t \gamma_{ij} &=& -2 \alpha K_{ij} + D_i \beta_j
   + D_j \beta_i\ , \label{dgdt}\\
\partial_t K_{ij} &=& \alpha (R_{ij} - 2 K_{ik} K^k_{~j}
 + K K_{ij})
- D_i D_j \alpha \label{dKdt} \\
&+& \beta^k \partial_k K_{ij} + K_{ik} \partial_j \beta^k
 + K_{kj} \partial_i \beta^k 
 - 8 \pi \alpha (S_{ij} - \frac{1}{2} \gamma_{ij} (S - \rho))\ , \nonumber
\end{eqnarray}
where $S$ and $S_{ij}$ are additional matter source terms.
The evolution equations guarantee that the field equations will 
automatically  satisfy the constraints on all future time slices 
identically, provided they satisfy them on the 
initial slice. Of course, this statement applies to the analytic
set of equations and not necessarily to their numerical counterparts.

Note that the $3+1$ formalism prescribes no equations 
for $\alpha$ and $\beta^i$. These four functions embody the four-fold
gauge (coordinate) freedom inherent in general relativity. 
Choosing them judiciously, especially in the presence of black holes,
is one of the main challenges of numerical relativity.

\subsection{The BSSN Scheme}

During the past decade, significant improvement in our ability to 
numerically integrate Einstein's equations stably in full $3+1$ dimensions 
has been achieved by recasting the original ADM system of equations. 
One such reformulation is the so-called 
BSSN scheme.\cite{rf:bssn} In this scheme, the physical metric and extrinsic
curvature variables are replaced in favor of the conformal metric and 
extrinsic curvature, in the spirit of 
the ``York-Lichnerowicz'' split:\cite{rf:yl}
\begin{eqnarray}
\tilde{\gamma}_{ij} &=& e^{-4\phi} \gamma_{ij},
~~~{\rm where}~~~ e^{4\phi} = \gamma^{1/3}\ ,\\
\tilde{A}_{ij} &=& \tilde{K}_{ij}- \frac{1}{3}\tilde{\gamma}_{ij} K\ .
\end{eqnarray}
Here a tilde $\tilde{}$ denotes a conformal quantity and $\gamma$ is the 
determinant of $\gamma_{ij}$.
At the same time, a connection function $\tilde{\Gamma}^i$ is introduced
according to
\begin{eqnarray} \label{Gamma}
\tilde{\Gamma}^i \equiv \tilde{\gamma}^{jk}\tilde{\Gamma}^i{}_{jk}
   = - \partial_j \tilde{\gamma}^{ij}\ .
\end{eqnarray}
The quantities that are independently evolved in this scheme are now
$\tilde{\gamma}_{ij}, \tilde{A}_{ij}, \phi, K$ and $\tilde{\Gamma}^i$.
The advantage is that the Riemann operator appearing in 
the evolution equations (cf. eqn.~(\ref{dKdt})) takes 
on the form,
\begin{eqnarray}
\tilde{R}_{ij} =
-\frac{1}{2} \underbrace{\tilde{\gamma}^{lm}\partial_m \partial_l
\tilde{\gamma}_{ij}}_{`Laplacian`} +
\underbrace{\tilde{\gamma}_{k(i}\partial_{j)}\tilde{\Gamma}^k}_
{\rm remaining\ 2nd\ derivs} + \cdots\ .
\end{eqnarray}
Thus the principal part of this operator, 
$\tilde{\gamma}^{lm}\partial_m \partial_l \tilde{\gamma}_{ij}$ is that of a
Laplace operator acting on the components of the metric $\tilde {\gamma}_{ij}$.
All the other second derivatives of the metric have been absorbed in the 
derivatives of the connection functions. The coupled evolution equations
for $\tilde{\gamma}_{ij}$ and $\tilde{A}_{ij}$
 (cf. eqns.~(\ref{dgdt}) and (\ref{dKdt}) then reduce essentially to a
wave equation,
\begin{eqnarray}
\partial_t^2 \tilde{\gamma}_{ij}
\sim \partial_t \tilde{A}_{ij}
\sim \tilde{R}_{ij} \sim  \nabla^2  \tilde{\gamma}_{ij}\ .
\end{eqnarray}
Wave equations not only reflect the hyperbolic nature of general relativity,
but can be implemented numerically in a straight-forward and stable manner.
By now, numerous simulations 
have demonstrated the
dramatically improved stability achieved in the BSSN scheme
over the standard ADM equations,
and considerable effort has gone into explaining the
improvement on theoretical grounds [see Ref.~\citen{rf:bs} for
references]. 

\section{Code Testing}

Much of the life of a computational relativist is 
devoted to code verification. This essential aspect of numerical
relativity combines physical and mathematical insight with dedication and 
fortitude. Code verification takes on diverse forms. 
For example, one component consists of
{\it convergence
testing} to check that as the numerical spacetime lattice is
shrunk to smaller scale to achieve higher resolution, the
integrations converge to a unique solution in a manner
consistent with the order of truncation of the numerical scheme.
Another component involves either identifying, or 
deriving from scratch, exact solutions, and then
{\it reproducing exact solutions numerically}.
In the case of vacuum spacetimes such test-bed solutions might 
consist of linearized, propagating gravitational waves and
stationary black holes. In the case of spacetimes containing 
hydrodynamic fluid sources, typical test-bed solutions consist of
shocks (e.g., the relativistic Riemann problem),
nonrotating and rotating stars in 
stable equilibrium, and spherical, homogeneous 
dust-ball collapse to a black hole (Oppenheimer-Synder collapse) expressed 
in various gauges. 
For spacetimes involving general relativistic magnetohydrodynamics, 
MHD wave propagation and MHD 
shocks, and
the excitation of MHD waves by linearized gravitational waves,
all yield exact solutions that serve as useful debugging tools.

{\it Monitoring constraint violation} is another mandatory aspect
of code validation for a dynamical simulation.
Not only must the fundamental 
Hamiltonian and momentum constraints, eqns.~(\ref{ham}) and
(\ref{mom}), be satisfied to some specified numerical tolerance at all times 
during a simulation, 
but 
any additional constraints associated with the adopted 
formalism must also be obeyed. In BBSN, for example, the quantities 
$\tilde{\gamma}_{ij}$ and $\tilde{\Gamma}^i$ are separately 
evolved, hence
the definition~(\ref{Gamma}) emerges as a constraint. For GRMHD simulations, it is
crucial to satisfy the magnetic field constraint equation, $D_i B^i = 0$.

Finally,  
{\it monitoring global conserved quantities} during 
a dynamical simulation provides an additional diagnostic tool.
Useful diagnostics include
the total (ADM) mass-energy $M$, the linear momentum $P^i$ and the 
angular momentum $J^i$ of the system,
allowing for any energy and momentum
carried off the computational grid by an outgoing 
flux of gravitational waves, 
matter and other forms of radiation. Other useful conserved quantities 
include the rest-mass $M_0$ when matter is present, and diagnostics like
fluid circulation ${\mathcal C}$ in adiabatic hydrodynamic 
flow (Kelvin's theorem). 

\section{Binary Neutron Star Mergers and Hypermassive Stars}
\label{bns}

The protagonist of several different astrophysical 
scenarios probed by recent numerical simulations is
a hypermassive star, typically a hypermassive neutron star. A hypermassive
star is an equilibrium fluid configuration that supports itself against 
gravitational collapse by {\it differential} rotation. Uniform rotation
can increase the maximum mass of a nonrotating, spherical
equilibrium (TOV) star by at most $\sim 20\%$, but differential rotation
can achieve a much higher increase.\cite{rf:sbs, rf:mbs}. Dynamical simulations
demonstrate\cite{rf:bss} that hypermassive stars can be 
constructed that are {\it dynamically} stable, provided 
the ratio of rotational kinetic to gravitational potential energy, $\beta$,
is not too large; for $\beta \ga 0.24$ the configuration is subject
to a nonaxisymmetric dynamical bar instability\cite{rf:sbs, rf:ssbs}.
However, all hypermassive stars are {\it secularly} unstable to
the redistribution of angular angular momentum by viscosity, magnetic
braking or turbulence, or any other agent that dissipates 
internal shear. Such a 
redistribution tends to drive a hypermassive
star to uniform rotation, which cannot
support the mass against collapse. Hence hypermassive stars are 
transient phenomena. Their formation
following, for example, the merger of the neutron stars in
a binary, or core collapse in a massive, rotating star, may ultimately 
lead to a `delayed' collapse to a black hole on secular (dissipative) 
timescales.  Such a collapse will be accompanied inevitably 
by a delayed gravitional wave burst.\cite{rf:bss}

The above scenario has become very relevant in light of the most
recent and detailed simulations of binary neutron star mergers in full general relativity. 
Dynamical simulations begin with initial data constructed to approximate 
astrophysically realistic binaries in quasistationary circular orbits at 
close separation. To construct such data, 
the initial value (constraint) equations (~\ref{dgdt}) and (~\ref{dKdt}) 
are solved for the
metric via the conformal thin-sandwich approximation,\cite{rf:cts} where
quasistationarity in the rotating frame of the binary is imposed by
assuming that the spacetime is endowed with a helical Killing vector.
The initial spatial metric is often assumed to be conformally
flat.  The initial matter profile is obtained by integrating
the Euler equation together with the constraints.
Two extreme opposite limits 
account for the stellar spins: 
corotational\cite{rf:cor} and irrotational
binaries\cite{rf:irr}. Irrotational binary stars are physically more realistic for
BNSs, since there is insufficient time for viscosity to achieve
corotation prior to merger.\cite{rf:bck}. 

\begin{figure}
   \begin{center}
   \includegraphics[height=5.0cm]{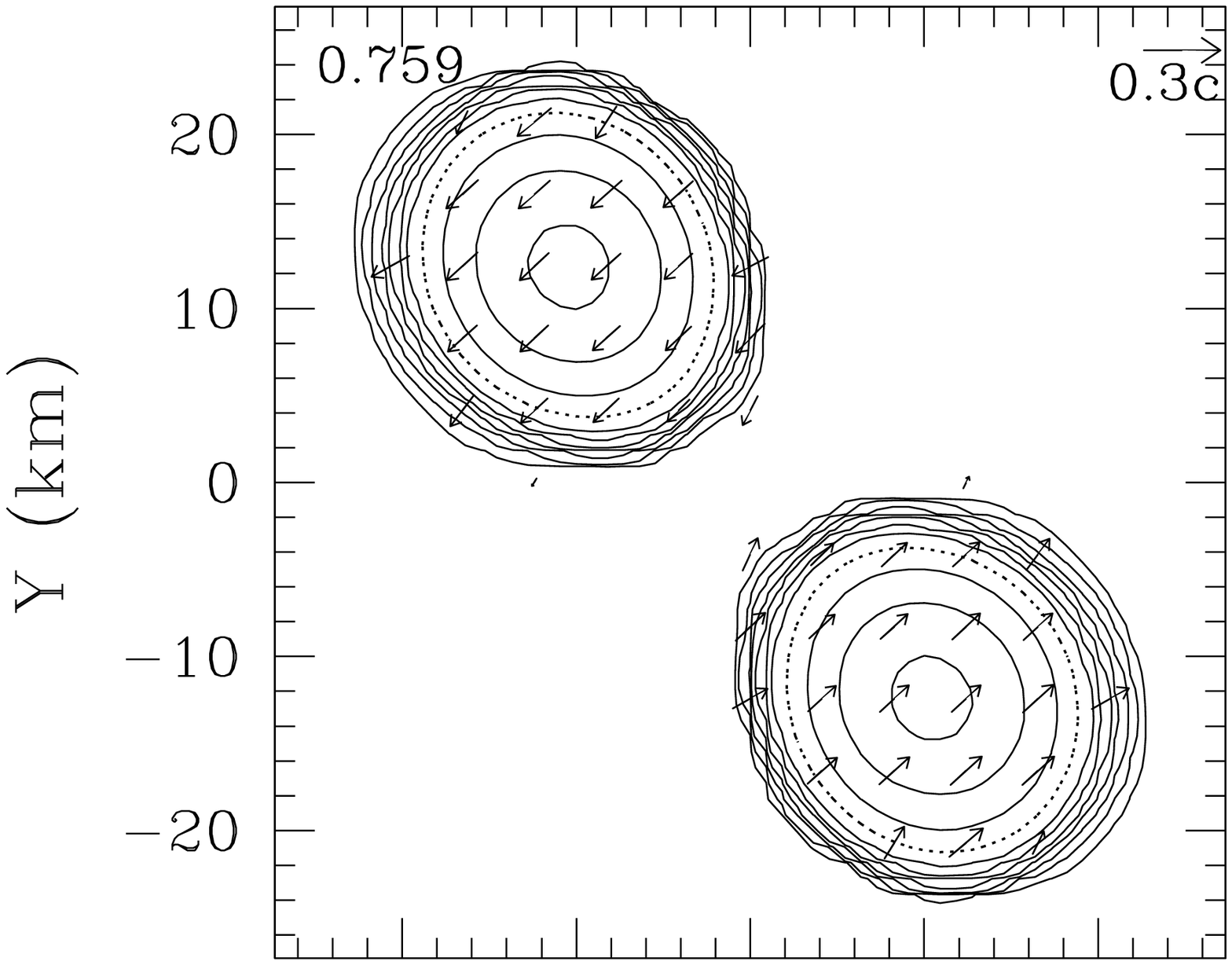}
   \includegraphics[height=5.0cm]{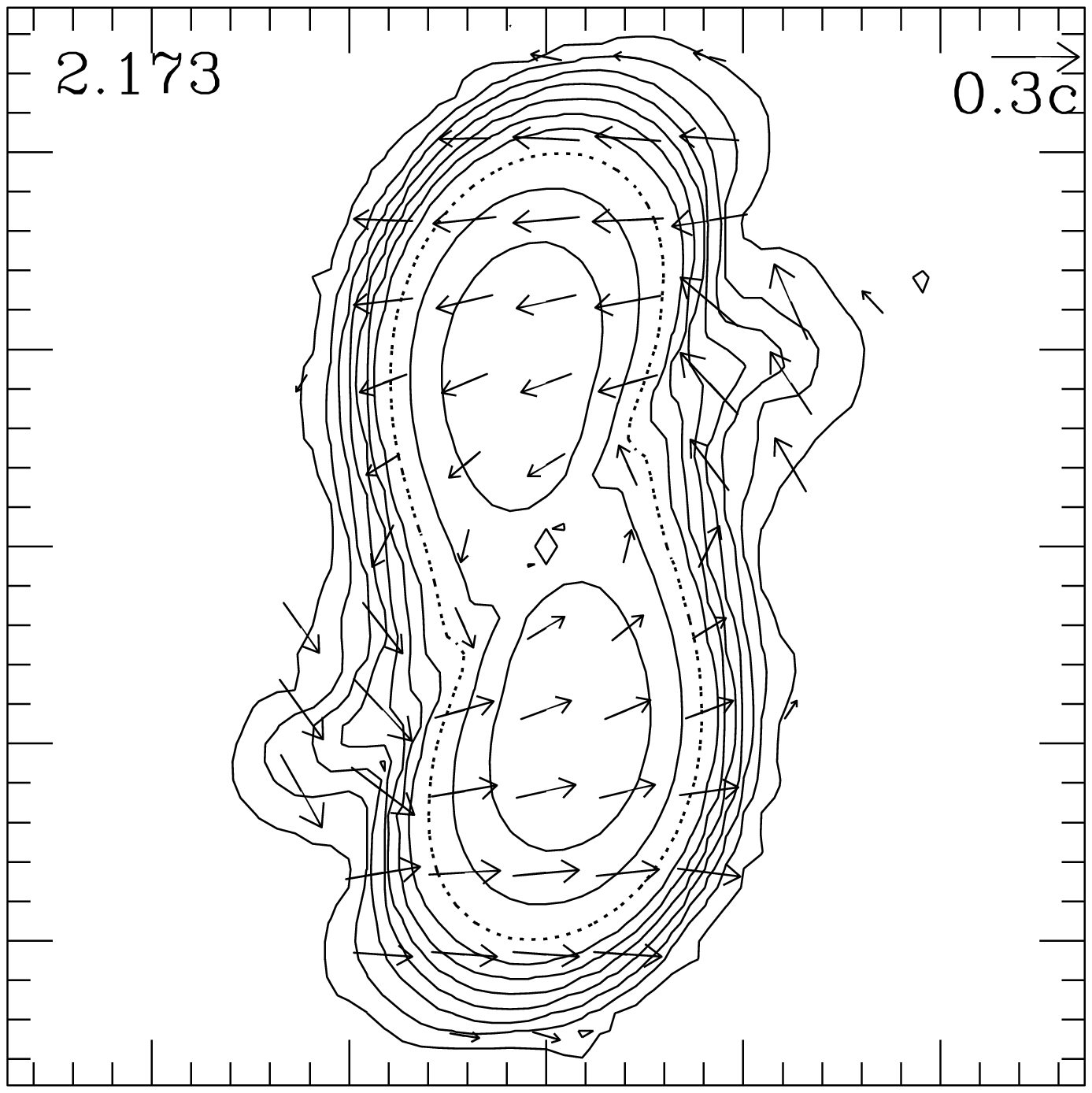}
   \includegraphics[height=5.0cm]{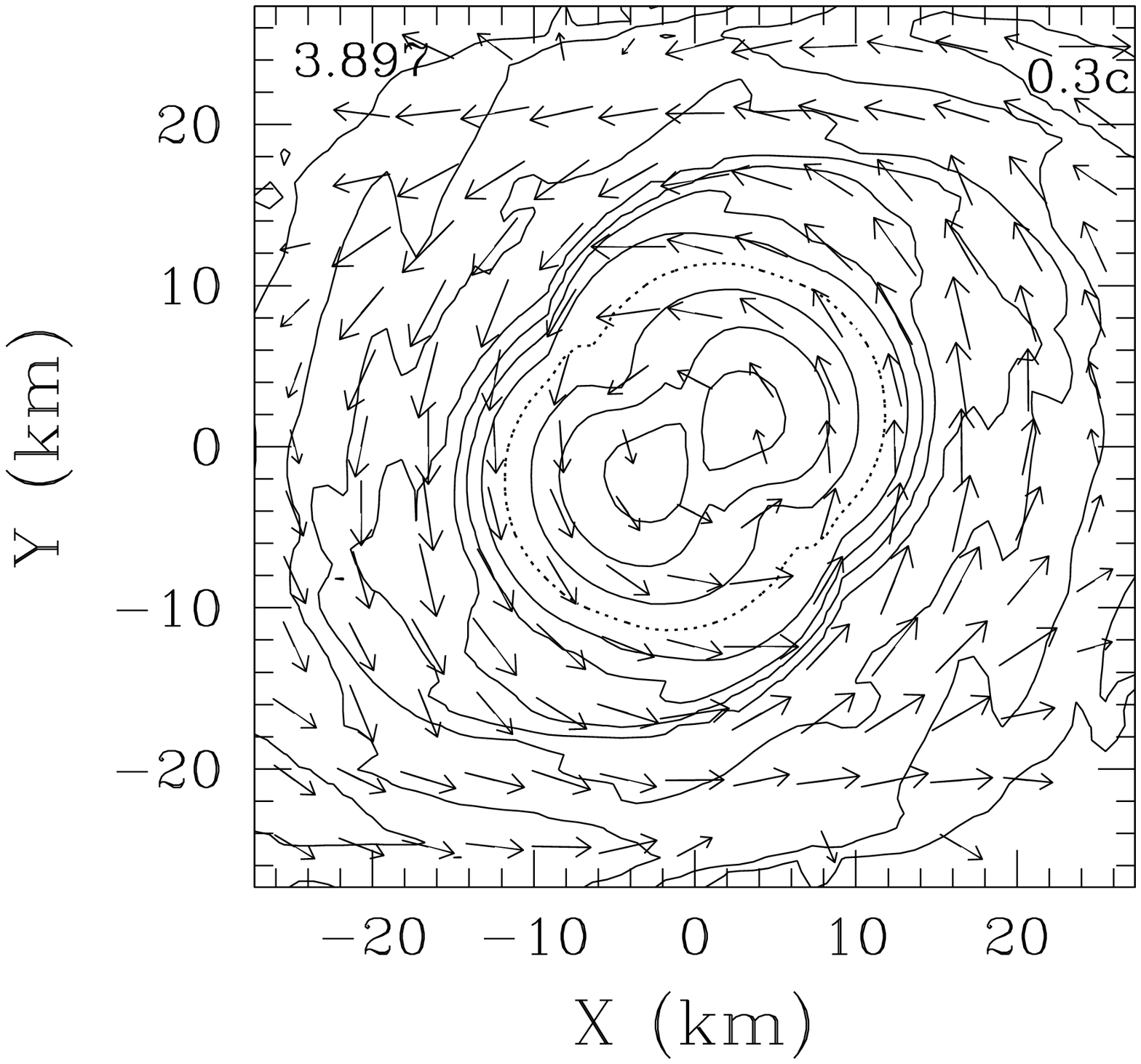}
   \includegraphics[height=5.0cm]{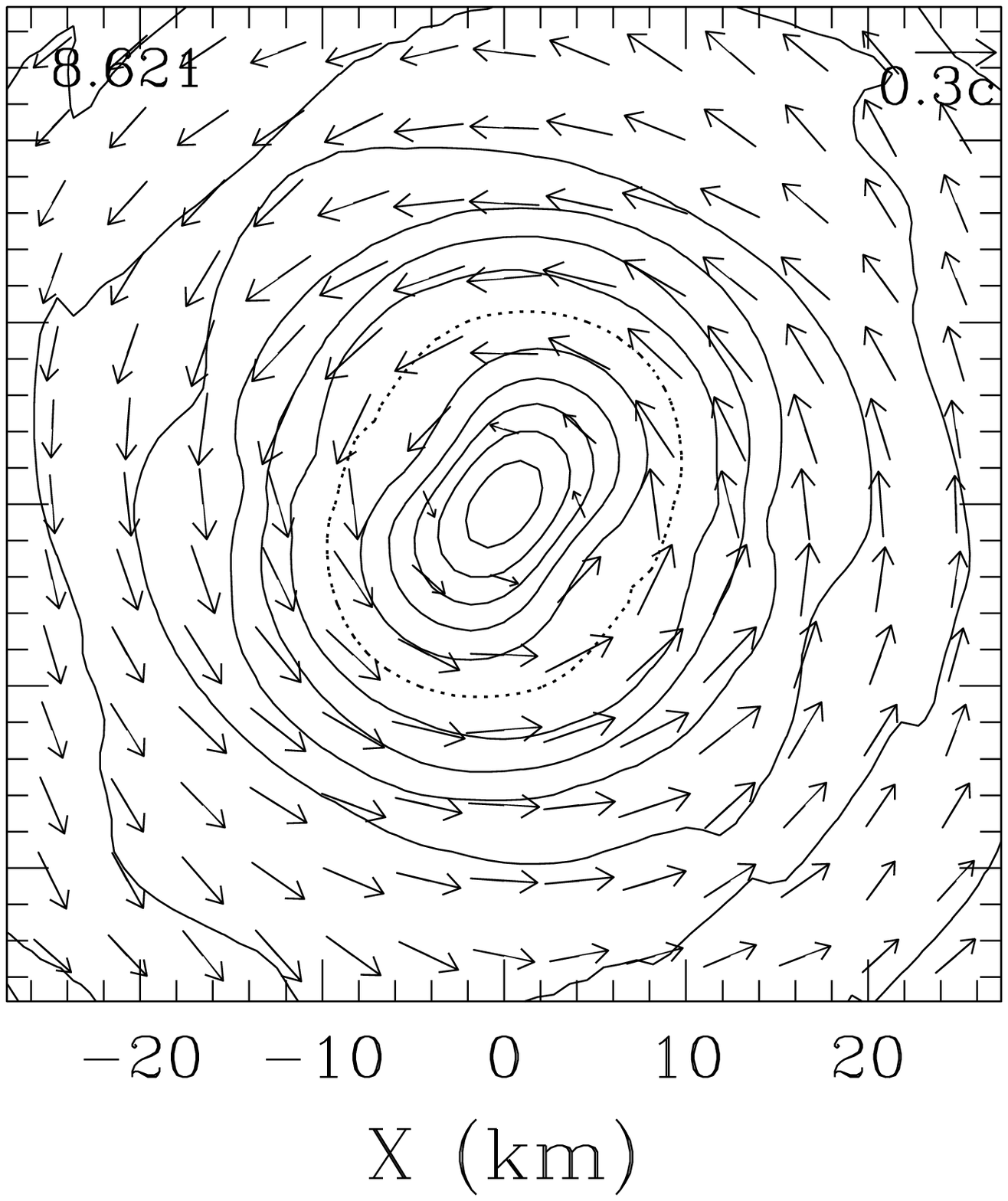}
   \caption{Formation of triaxial hypermassive remnant following 
   BNS merger in $2.7 M_{\odot}$ system. Snapshots of density contours 
   are shown in the equatorial plane.
   The number in the upper left-hand corner denotes the elapsed time 
   in ms; the initial orbital period is 2.11 ms. Vectors indicate 
   the local velocity field. 
   [From Shibata, Taniguchi and Uryu 2005.]}
   \label{fig2}
   \end{center}
   \end{figure}

State-of-the-art, fully relativistic simulations of BNSs have been performed 
by Shibata and his collaborators. They consider mergers of $n=1$ polytropes
\cite{rf:stu03}, as well as configurations obeying a realistic 
nuclear equation of 
state (EOS)\cite{rf:stu05}. They treat mass ratios $Q_M$ in the 
range $0.9 \leq Q_M \leq 1$, consistent with the range of $Q_M$ in 
observed binary pulsars with accurately determined masses\cite{rf:tc99}. 
The key 
result is that there exists a critical mass 
$M_{\rm crit} \sim 2.5 - 2.7 M_{\odot}$ of the binary system above which the
merger leads to prompt collapse to a black hole, and below which the merger 
forms a hypermassive remnant. 
The hypermassive remnant undergoes delayed
collapse in about $\sim 100$ ms and emits a delayed gravitational wave burst.
Most interesting, prior to collapse, the remnant forms a triaxial 
bar when a realistic EOS is adopted (see Fig.~\ref{fig2}) and the bar 
emits quasiperiodic gravitational waves at a frequency $f \sim 3 - 4$ kHz.
Such a signal may be detectable by Advanced LIGO. 
It is interesting that for the adopted EOS, the mass $M_{\rm crit}$ is close
to the value of the total mass found in each of the observed binary pulsars.
Given that the
masses of the individual stars in a binary
can be determined by measuring the gravitational wave signal emitted 
during the adiabatic, inspiral epoch prior to plunge and merger,
the detection (or absence) of any
quasiperiodic emission from the hypermassive remnant prior to delayed 
collapse may significantly constrain models of the nuclear EOS. 

The possibility of a hypermassive neutron star remnant formed 
following a BNS merger had been foreshadowed in earlier Newtonian 
simulations\cite{rf:newt}, post-Newtonian
simulations\cite{rf:pnewt} and conformally flat general relativistic 
simulations\cite{rf:cfgrt}. However, the recent fully  relativistic 
simulations reported in Ref.~\citen{rf:stu05} provide the strongest 
theoretical evidence to date, although the details undoubtedly 
depend on the adopted EOS.  Triaxial equilibria can arise only in 
stars that can support sufficiently high values of $\beta$  
exceeding the classical bifurcation point at $\beta \approx 0.14$; 
reaching such high values
requires  EOSs with adiabatic indicies exceeding $\Gamma \approx  2.25$ in 
Newtonian configurations, and
comparable values in relativistic stars. 
It is not yet known whether the true nuclear EOS in neutron stars 
is this stiff, or what agent for redistributing 
angular momentum in a hypermassive star dominates
(e.g., magnetic fields, turbulent viscosity or gravitational radiation).
But it is already evident that hypermassive stars are likely to form 
from some mergers and that they will survive many dynamical
timescales before undergoing delayed collapse. The recent 
measurement\cite{rf:nice}
of the mass of a neutron star in a NS-white dwarf binary of 
$M_{\odot} = 2.1$ establishes an observational lower limit for 
the maximum mass of a neutron star; such a high value suggests that mergers in
typical BNSs may form hypermassive stars more often than undergoing 
prompt collapse.

\section{Black Hole Excision}

The challenge of simulating a black hole numerically is 
avoiding encountering the spacetime singularity inside the hole. Failure to
do so results in infinities that cause the numerical integrations to terminate.
While suitable gauge conditions can postpone the appearance of singularities,
they are susceptible to ``grid stretching'' inaccuracies that usually infect the
metric near the horizon and grow with time; see Ref.~\citen{rf:bs} for 
discussion and Fig.~\ref{fig7} in Section~\ref{GRMHD} for an example.
One of the most promising new methods of dealing with black hole
singularities is black hole excision.  This method, first suggested
by Unruh~\cite{rf:u87}, exploits the fact that the singularity resides
inside an event horizon,
a region that is casually disconnected from the rest of the
universe.  Since no physical information propagates from inside the
event horizon to outside, one should be able to evolve the exterior
independent of the interior spacetime.  Inside the event horizon,
causality entitles us to do almost
``anything'' which will produce a stable exterior evolution.  
In particular, one
can excise the computational domain inside the horizon 
and replace it with suitable boundary conditions at its outer surface.

Although it is guaranteed that no physical signal can propagate from
inside the horizon to outside, unphysical signals often can propagate
in evolution codes.  Gauge
modes can move acausally for many gauge conditions.  Although they
carry no physical content, such modes may destabilize the code.  Thus,
the choice of gauge is crucial to obtaining good post-excision evolutions. 
In addition, constraint-violating modes can, for some formulations of
the field equations, propagate acausally, creating inaccuracies and
instabilities.  Thus, the choice of formulation is also crucial to
obtaining good post-excision evolutions. 

The feasibility of black hole excision was first demonstrated in
spherically symmetric 1+1 dimensional evolutions of a single black
hole in the presence of a self-gravitating scalar field
\cite{rf:ss,rf:scf}.
Excision was also implemented early on to study the spherically
symmetric collapse of collisionless matter to a black hole in Brans-Dicke
theory~\cite{rf:sst95}.
Full $3+1$ evolutions of vacuum black hole spacetimes were attempted 
with excision using the standard ADM formulation, both for
a stationary~\cite{rf:amsst95} and for a boosted black hole~\cite{rf:BBHGCA98}.
Although the introduction of excision improved the behavior of these black
hole simulations, long-term stability could not be achieved due to
instabilities endemic to the unmodified ADM formulation.

Using excision in the
BSSN formulation, several groups
have evolved stationary, vacuum black hole spacetimes with and without spin
for arbitrarily long times~\cite{rf:exbssn}.  Long-term
stability has also been achieved using hyperbolic formulations of
the field equations~\cite{rf:exhyp} and 
characteristic formulations~\cite{rf:exchr}.
Excision has also been
used to simulate the grazing collision of two black holes~\cite{rf:b00}
and binary black holes initially in circular orbit and remaining in orbit 
for approximately one orbital period~\cite{rf:btj04}.

Some of the most recent advances involving black hole excision involve its
incorporation in $3+1$ BSSN relativistic hydrodynamics schemes
that treat perfect gases\cite{rf:exhydro}, imperfect gases with 
viscosity\cite{rf:dlss}
and GRMHD fluids\cite{rf:dlss05,rf:ss05}. In addition, black hole 
excision has been implemented in a new $3+1$ scalar wave scheme in 
generalized harmonic coordinates
that has been used to follow the plunge, merger and ringdown of a binary
black hole system formed from the collapse of scalar waves\cite{rf:pre}. 

\section{Relativistic Hydrodynamics With Viscosity}

The advent of robust, new formulations of the $3+1$ equations, like
BSSN, now makes it possible to track the {\it secular} evolution of a 
relativistic star over many dynamical timescales. 
The implementation of black hole excision
makes it possible to track the collapse of a radially unstable star for 
a time interval $\Delta t \gg M$ following the formation of a black hole.
Both of these advances were recently exploited to follow 
the long-term evolution of a hypermassive star driven radially unstable 
to collapse by the action of shear viscosity\cite{rf:dlss}. This treatment 
was essentially the first time that
the tools of numerical relativity were employed to integrate the 
relativistic $3+1$ Navier-Stokes equations for an imperfect gas. 
The stress-energy 
tensor for an imperfect fluid containing viscosity is 
$T^{ab}=T^{ab}_{\rm ideal} + T^{ab}_{\rm visc}$,
where $T^{ab}_{\rm ideal}$ is the familiar expression for an ideal gas
and $T^{ab}_{\rm visc}$ is given by~\cite{rf:mtw}
\begin{eqnarray} \label{Tvis}
T^{ab}_{\rm visc} &=& -2 \eta \sigma^{ab} - \zeta\theta P^{ab}\ .
\end{eqnarray}
In eqn.~(\ref{Tvis}), $\eta$ is the shear viscosity, 
$\sigma^{ab}$ is the shear tensor,   
$\zeta$ is the bulk viscosity, $\theta$ is the expansion and 
$P^{ab}= g^{ab} + u^a u^b$ is the
projection tensor of the fluid, where $u^a$ is the fluid 4-velocity. 
In the numerical simulations of hypermassive stars,
the shear viscosity was chosen to be proportional to the gas pressure
and the bulk viscosity was set equal to zero, since it is the shear
viscosity that redistributes angular momentum.
The magnitude of $\eta$, though enhanced, was chosen to maintain the 
inequality $t_{\rm visc} \gg t_{\rm dyn}$ 
between the viscous and dynamical timescales, thereby satisfying a
physically realistic, but computationally challenging, timescale hierarchy.
The viscosity heats up the fluid according to 
\begin{equation}
\rho_0 T (ds/d\tau)_{\rm visc} = 2\eta\sigma^{ab}\sigma_{ab}\ .
\end{equation}
In Ref.~\citen{rf:dlss}, two extreme opposite cases are treated
for the net cooling of the fluid 
(e.g, by electromagnetic radiation or
neutrino emission): ``no cooling'', appropriate whenever
$t_{\rm cool} \gg  t_{\rm visc}$, where $t_{\rm cool}$ is the thermal
cooling timescale, and ``rapid cooling'', appropriate
when the inequality is reversed. In axisymmetry, the evolution can be
tracked over dozens of rotation periods (thousands of $M$).

The key result is that viscosity operating in a hypermassive star 
generically leads to the formation of a compact, uniformly rotating
core surrounded by a low-density disk (``viscous braking of differential
rotation'').  The core typically  grows more massive in time, ultimately 
becoming unstable to gravitational collapse. The simulations can track 
the secular evolution of the star, its dynamical collapse to a black hole,
and, using black hole excision, relaxation to a final, stationary, 
black hole-disk quasiequilibrium state. The  
masses and spins of the final black hole and ambient disk are 
in good agreement with the values inferred from an analytic estimate 
based on the stellar density and angular momentum profiles at the onset of 
collapse.\cite{rf:ss02}. 

One interesting finding is that, for a given viscosity law, the
secular evolution exhibits a simple scaling behavior provided
$t_{\rm visc} \gg t_{\rm dyn}$: 
the evolving matter profiles are identical, independent of the magnitude
of the viscosity, except that the timescale for the matter to reach a
certain state prior to collapse varies inversely with the viscosity, i.e., 
$t_2 = t_1 (\eta_1/\eta_2)$.

\begin{figure}
       \centerline{\includegraphics[height=10.0 cm]
                                   {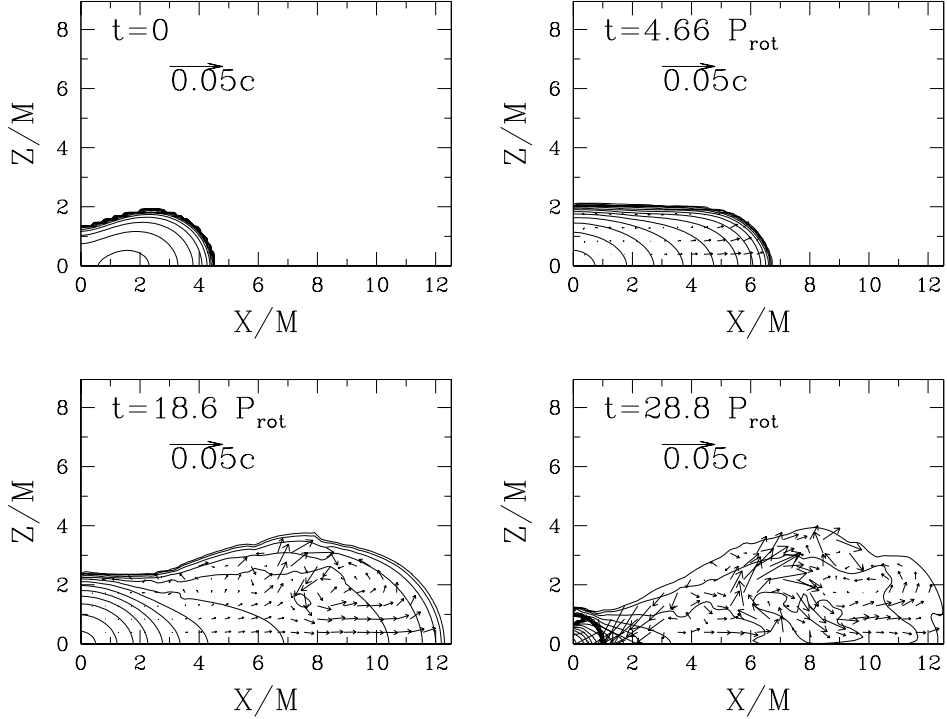}}
   \caption
     {Meridional rest-mass density contours and velocity field
      at select times (in units of central rotation period $P_{\rm rot}$) 
      for a hypermassive star with shear viscosity.
      The system is axisymmetric and experiences no cooling.
      The levels of the contours (from inward to outward) are 
      $\rho_0/\rho_{0,\rm max}= 10^{-0.15 (2 j +0.6)}$, 
      where $j=0,\ 1, \ ...\ 12$.  In the lower right panel 
      ($t=28.8 P_{\rm rot}$), the thick curve denotes the apparent horizon.
      [From Duez et al. 2004.]}
   \label{fig3}
   \end{figure}

Fig.~\ref{fig3} depicts the evolution without cooling of a hypermassive, $n=1$ 
polytrope  with $J/M^2 = 1$ that exceeds the maximum mass of a 
uniformly rotating polytrope with the same index
by a factor of 1.47. Viscosity drives secular evolution, leading to
catastrophic collapse and the formation of a rotating black hole containing
$77 \%$ of the total initial rest mass and $35 \%$ of the initial spin 
($J_h/M_h^2 \approx 0.5$) . The
remainder of the rest mass and spin go into the ambient disk, which 
continues to accrete slowly onto the black hole. Because the ambient disk has 
appreciable mass, the spacetime is not strictly Kerr. However, an
ergoregion develops about hole, as well as an 
ISCO (innermost stable circular orbit); the ISCO 
is evident by tracking the gas flow in the equatorial plane near the 
black hole and noting the transition from secular inspiral to dynamical
plunge.

\section{Binary Black Hole -- Neutron Stars}

The first treatments of both the initial data and the evolution of
BBHNSs that construct the combined BH-NS spacetime in a general
relativistic framework have been presented recently. The initial data
in these treatments consists of two stars in circular orbit and is 
constructed in the conformal thin-sandwich approximation by assuming 
that the spacetime possesses an approximate helical Killing vector.
This condition is designed to yield a nearly stationary metric in the
frame corotating with the orbit. Current implementations
treat the regime $M_{\rm NS}/M_{\rm BH} \ll 1$.\cite{rf:bhns} 
Two distinct cases are considered for the
conformally related background metric ${\tilde \gamma}_{ij}$:
a Kerr-Schild BH metric and an isotropic (conformally flat) BH metric. 
Two opposite limiting cases are considered for the NS spin, corotation 
and irrotation, for several different polytropic EOSs.
The two background metrics 
give comparable values for the orbital
angular velocity and the maximum stellar density 
at the Roche (tidal break-up) limit for the binaries. Any difference in 
these scalar invariants 
reflects the degree to which the physical models 
are sensitive to the assumed background metric; evolution calculations
are required to interpret the differences.

\begin{figure}
       \centerline{\includegraphics[height=10.0 cm]
                                   {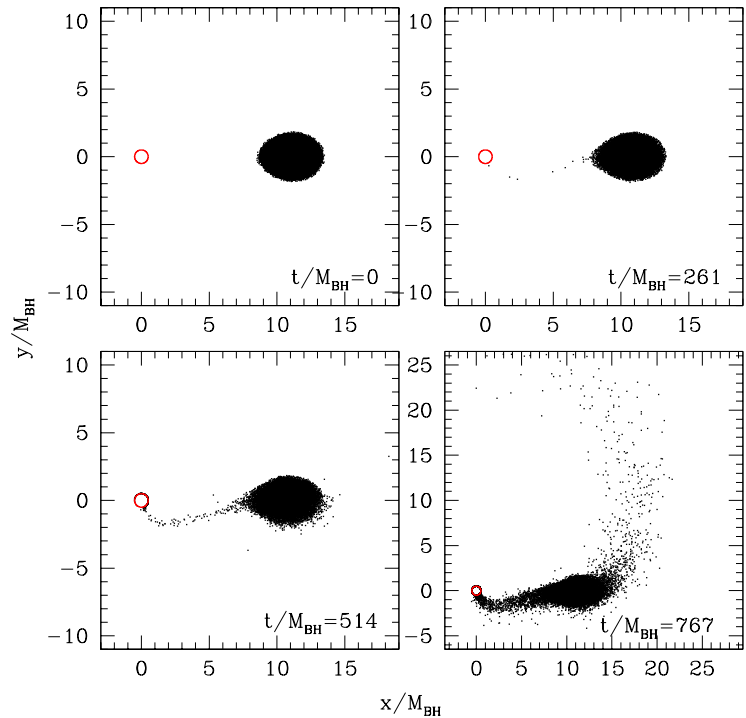}}
   \caption
     {Snapshots of SPH particle configurations projected into the 
  orbital plane at select times corresponding to the initial configuration and
  $1$, $2$, and $3$ full orbits, respectively. 
  The BH horizon is shown as a circle at an isotropic radius $r=0.5M_{\rm BH}$
  The binary separation begins just
  outside the Roche tidal radius.
  Radiation-reaction initially drives an
  inwardly directed mass-transfer stream onto the BH, until somewhere
  after $t/M_{\rm BH}=700$, when the expansion of the NS becomes unstable and
  tidal disruption occurs. [From Faber et al. 2005.]}
   \label{fig4}
   \end{figure}

Evolution calculations have been performed with the corotational, 
conformally flat, initial data, with the approximation that the spatial metric 
remains conformally flat.\cite{rf:fab05} The matter is evolved using a
relativistic Lagrangian smoothed partical hyrodynamics (SPH) scheme.
The first simulations have been for neutron stars with
low-compaction, $M_{\rm NS}/R_{\rm NS} = 0.04$, with 
$M_{\rm NS}/M_{\rm BH} = 0.1$. A radiation-reaction 
potential is
included in the Euler equation to trigger the inward spiral
(see Fig.~\ref{fig4}).

These preliminary simulations show that
mass transfer, when it begins while the neutron star orbit is still
outside the BH ISCO, is more unstable than is
typically predicted by analytical formalisms, i.e., 
$d M_{\rm NS}/dt \gg M_{\rm NS}/ t_{GW}$, where $t_{\rm GW}$ is the 
gravitational wave inspiral timescale. This dynamical mass
loss is found to be the driving force in determining the subsequent
evolution of the binary orbit and the neutron star, which typically
disrupts completely within a few orbital periods.  The majority of the
mass transferred onto the black hole is accreted promptly; a
significant fraction ($\sim 30\%$) of the mass is shed outward as well, some of
which will become gravitationally unbound and ejected completely from
the system.  The remaining portion forms an accretion disk around the
black hole, and could provide the energy source for short-duration
gamma ray bursts. More detailed studies are underway.

\section{General Relativistic MHD}
\label{GRMHD}

Magnetic fields play a crucial role in determining the evolution of
many relativistic objects.  In any highly conducting astrophysical
plasma, a frozen-in magnetic field can be amplified appreciably by gas
compression or shear.  Even when an initial seed field is weak,
the field can grow in the course of time to significantly influence the gas
dynamical behavior of the system.  In problems where the self-gravity of
the gas can be ignored, simulations can be performed in a fixed background
spacetime.  Some accretion problems fall 
into this category. In
many other problems, the effect of the magnetized fluid on the metric
cannot be ignored, and the two must be evolved self-consistently.
The final fate of many of these relativistic astrophysical systems,
and their distinguishing observational signatures, may hinge on the role
that magnetic fields play during the evolution. Some of these
systems are promising sources of gravitational radiation,
while others also may be responsible for gamma-ray bursts. 

Examples of astrophysical scenarios involving
strong-field {\em dynamical} spacetimes in which MHD effects may play a
decisive role include the following:

$\bullet$ The merger of a BNS. The coalescence can lead to
the formation of a hypermassive star supported by differential
rotation, as discussed in Section \ref{bns}. 
The growth of magnetic fields through 
magetic braking (winding) and the magnetorotational instability (MRI) 
is alone sufficient to drive the star unstable,
even if the seed magnetic field is
small~\cite{rf:bss,rf:shap00}. This process can lead to delayed
collapse and massive disk formation, accompanied
by a delayed gravitational wave burst. 

$\bullet$ Core collapse in a supernova. Core collapse may induce
differential rotation, even if the rotation of the progenitor at the
onset of collapse is only moderately rapid and almost uniform. 
Differential rotation can wind up a frozen-in
magnetic field to high values, at which point it may provide a
significant source of stress, which could affect the explosion~\cite{rf:lw70}. 

$\bullet$ The generation of gamma-ray bursts (GRBs). Short-duration 
GRBs are thought to result from BNS
mergers~\cite{rf:npt,rf:kr98}, or tidal disruptions of neutron stars
by black holes~\cite{rf:rj}, or hypergiant flares of `magnetars' 
associated with the soft gamma-ray repeaters~\cite{rf:ngpf}. 
Long-duration GRBs likely result from
the collapse of rotating, massive stars which form
black holes (`collapsars')~\cite{rf:mw}.
In current scenarios, the burst is
powered by the extraction of rotational energy from the neutron star
or black hole, or from the remnant disk material formed around the black
hole~\cite{rf:vk}. Strong magnetic fields provide a likely
mechanism for extracting this energy on the required timescale and
driving collimated GRB outflows in the form of relativistic
jets~\cite{rf:mr}. Even if the initial magnetic
fields are weak, they can be amplified to the required values by
differential motions~\cite{rf:mri}.

$\bullet$ Supermassive black hole (SMBH) formation.  The cosmological origin
of the SMBHs observed in galaxies and quasars is one of the key unsolved
problems of structure formation in the early universe. Several hypotheses 
involve relativistic fluids in which
magnetic fields can play an important role.  It is thought that
SMBHs start from smaller initial seed black holes, which grow to
supermassive size by a combination of accretion and mergers.  The seed
black holes might be provided by the collapse of massive
($M\sim 10^2M_{\odot}$) Population III stars~\cite{rf:mr01}.  If so, magnetic
forces can affect the collapse leading to the formation of these seeds,
as well as their growth by accretion~\cite{rf:gsm}. In fact, 
for Eddington-limited disk accretion,
magnetic fields could be crucial to explaining how the growth of 
such seeds could be sufficiently rapid to account for the high
SMBH masses inferred for the very youngest quasars~\cite{rf:shap05}.
QSO SDSS 1148+5251, the quasar with the highest known 
redshift ($z=6.43$) and believed to be powered
by a SMBH that has grown to a  mass $M \sim 10^9M_{\odot}$, is a case in point.
An alternative possibility is that
SMBHs form directly from the catastrophic collapse of supermassive stars
(SMSs)~\cite{rf:r84}.  The evolution of a SMS 
may proceed differently, depending on
whether the SMS rotates uniformly, in which case it ultimately 
undergoes collapse to 
a SMBH\cite{rf:sshap02}, or differentially, in which case it 
becomes unstable to a bar mode
prior to reaching the onset of collapse~\cite{rf:ns}.
Magnetic fields and the turbulence they generate provide the
principle mechanisms that damp differential rotation in SMSs ~\cite{rf:zn}, and thus such fields may determine their ultimate fate.

$\bullet$ The r-mode instability in rotating neutron stars. This
instability has been proposed as a possible mechanism for
limiting the angular velocities in neutron stars and producing
observable quasi-periodic gravitational waves~\cite{rf:rmodes}.  However,
preliminary calculations suggest that if
the stellar magnetic field is strong enough, r-mode oscillations will
not occur~\cite{rf:rlms}. Even if the initial field is weak,
fluid motions produced
by these oscillations may amplify the magnetic field and eventually
distort or suppress the r-modes altogether. (R-modes may also be
suppressed by non-linear mode coupling~\cite{rf:saftw} or
hyperon bulk viscosity~\cite{rf:jl}.)

\begin{figure}
       \centerline{\includegraphics[width=5 cm,angle=270]
                                   {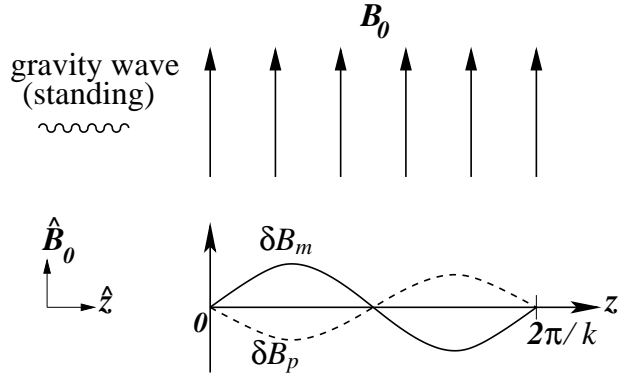}}
   \caption
     { 
Perturbation of the magnetic field $\delta \ve{B}=(\delta B_m
+ \delta B_p) \hat{\ve{B}}_0$ induced by a standing gravitational wave when
$\ve{k} \cdot \ve{v}_A=0$. The
unperturbed magnetic field $\ve{B}_0$ is perpendicular to the
wave vector of the gravitational wave $\ve{k}=k\hat{\ve{z}}$. Both stationary
modes $\delta B_p$ and $\delta B_m$ have the same $\sin \!kz$ spatial
dependence, but they oscillate with different
amplitudes and at different frequencies. [From Duez et al. 2005b]
}
   \label{fig5}
   \end{figure}

\begin{figure}
       \centerline{\includegraphics[height=8.0 cm]
                                   {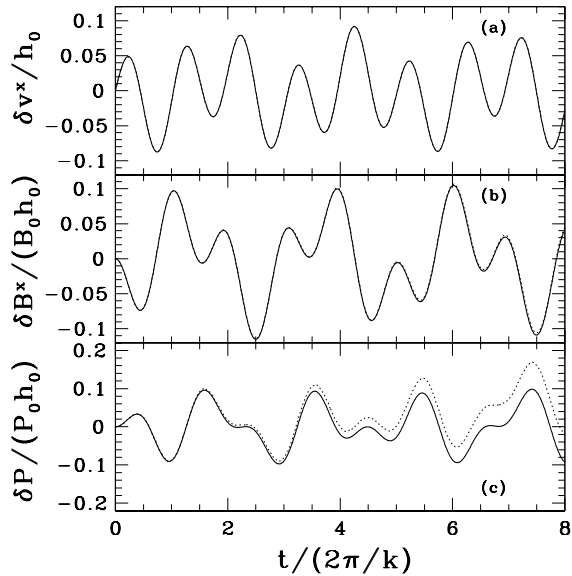}}
   \caption
     {
Analytic and numerical solutions for the perturbations of a
magnetized fluid due to the presence of a gravitational wave.
The thick solid and thin dotted lines
represent, respectively, the analytic and numerical solutions, though the
two lines are not readily distinguishable in plots (a) and (b).  All quantities
are evaluated at $kz = \pi/4$ and are normalized as indicated.
Time is normalized by the gravitational wave period. The drift noticeable in
the pressure perturbation is not due to numerical error, but to nonlinear contributions
neglected in the analytic solution. [From Duez et al. 2005a.]
  }
   \label{fig6}
   \end{figure}

Several numerical codes which evolve the 
GRMHD equations on {\it fixed} Schwarzschild or Kerr black hole spacetimes have
been developed in the past decade~\cite{rf:fixed}.
These codes have been used to study the structure of accretion flows onto Kerr
black holes, the Blandford-Znajek effect in low-density
regions near the hole, and the formation of GRB jets.
Until very recently, little progress has been made to treat
GRMHD flows in {\it dynamical} spacetimes beyond the
earliest attempt thirty years ago by Wilson~\cite{rf:w75}, who simulated 
the collapse of a
rotating SMS with a frozen-in poloidal magnetic field in axisymmetry 
using a code that assumed the conformal flatness approximation for the 
spatial metric, thereby eliminating all
gravitational radiation. In the past year a new code has been developed
to evolve GRMHD fluids in dynamical spacetimes without 
approximation.~\cite{rf:dlss05} 
The code solves the Einstein-Maxwell-MHD
system of coupled equations both in axisymmetry and in 3+1 dimensions.
It evolves the metric by integrating the BSSN
equations, and uses a conservative shock-capturing scheme to evolve
the GRMHD equations.  The contribution of the electromagnetic field to the
stress energy tensor in the GRMHD limit may be simply expressed as
\begin{equation}
T^{ab}_{\rm em} = b^2 u^a u^b + \frac{1}{2}b^2 g^{ab} - b^a b^b\ ,
\end{equation}
where $u^a$ is the fluid 4-velocity, 
$b^a = B^a/\sqrt{4 \pi}$, $B^a = u_a F^{*ab}$ is the magnetic field measured
by an observer comoving with the fluid, and $F^{*ab}$ is the dual of the Faraday
tensor.
The code gives accurate results in standard GRMHD code-test
problems, including magnetized shocks and magnetized Bondi accretion flow.
To test the code's ability to evolve the GRMHD equations in a dynamical
spacetime, perturbations of a homogeneous, magnetized fluid (i.e.
Alfv\'{e}n waves, and both fast and slow magnetosonic waves)
excited by a gravitational plane wave were studied, and good agreement 
between the numerical and new analytic 
solutions were found (see Fig.~\ref{fig5} and
~\ref{fig6}). This code has been followed by other robust codes
that have been 
constructed in a similar fashion, also using the BSSN formalism.\cite{rf:ss05,rf:a05}.

\begin{figure}
       \centerline{\includegraphics[height=18.0 cm]
                                   {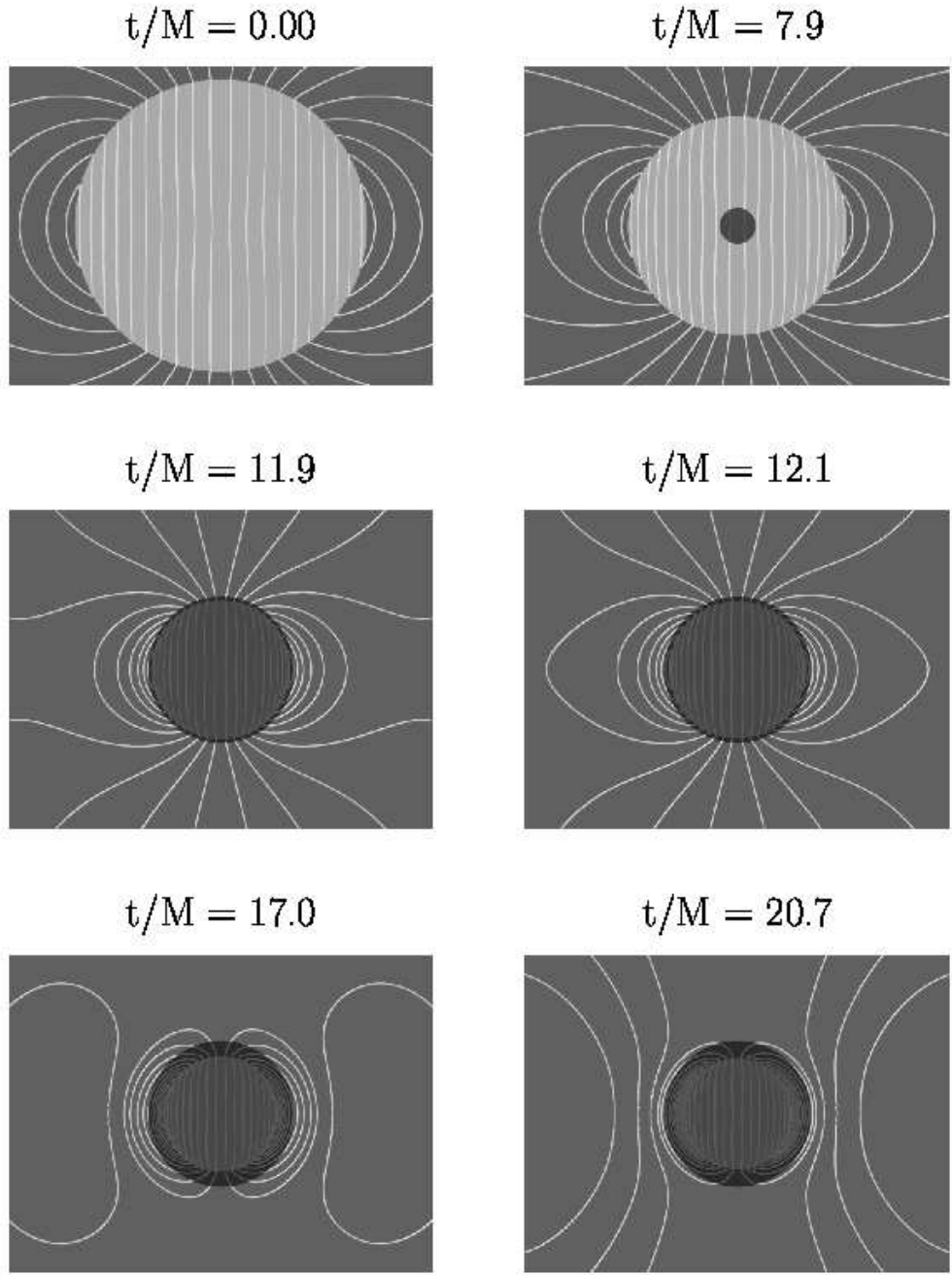}}
   \caption
     {
Snapshots of the interior and exterior magnetic field lines
on select {\it maximal} time slices. 
The star collapses from rest at an initial areal
radius $R_s(0) = 4 M$. 
The light shaded sphere covers the matter interior;
the black shaded sphere covers the region inside the event horizon.
In this gauge the stellar surface approaches a limit surface at
$r_s \approx 1.5 M$ at late times. The horizon grows to its final
value $r_s = 2M$ once all the matter crosses inside this radius.
Soon after the last snapshot at $t = 20.7$ grid stretching causes the
quality of the numerical integration to deteriorate in this gauge.
[From Baumgarte and Shapiro 2003b.]
  }
   \label{fig7}
   \end{figure}

\begin{figure}
       \centerline{\includegraphics[height=18.0 cm]
                                   {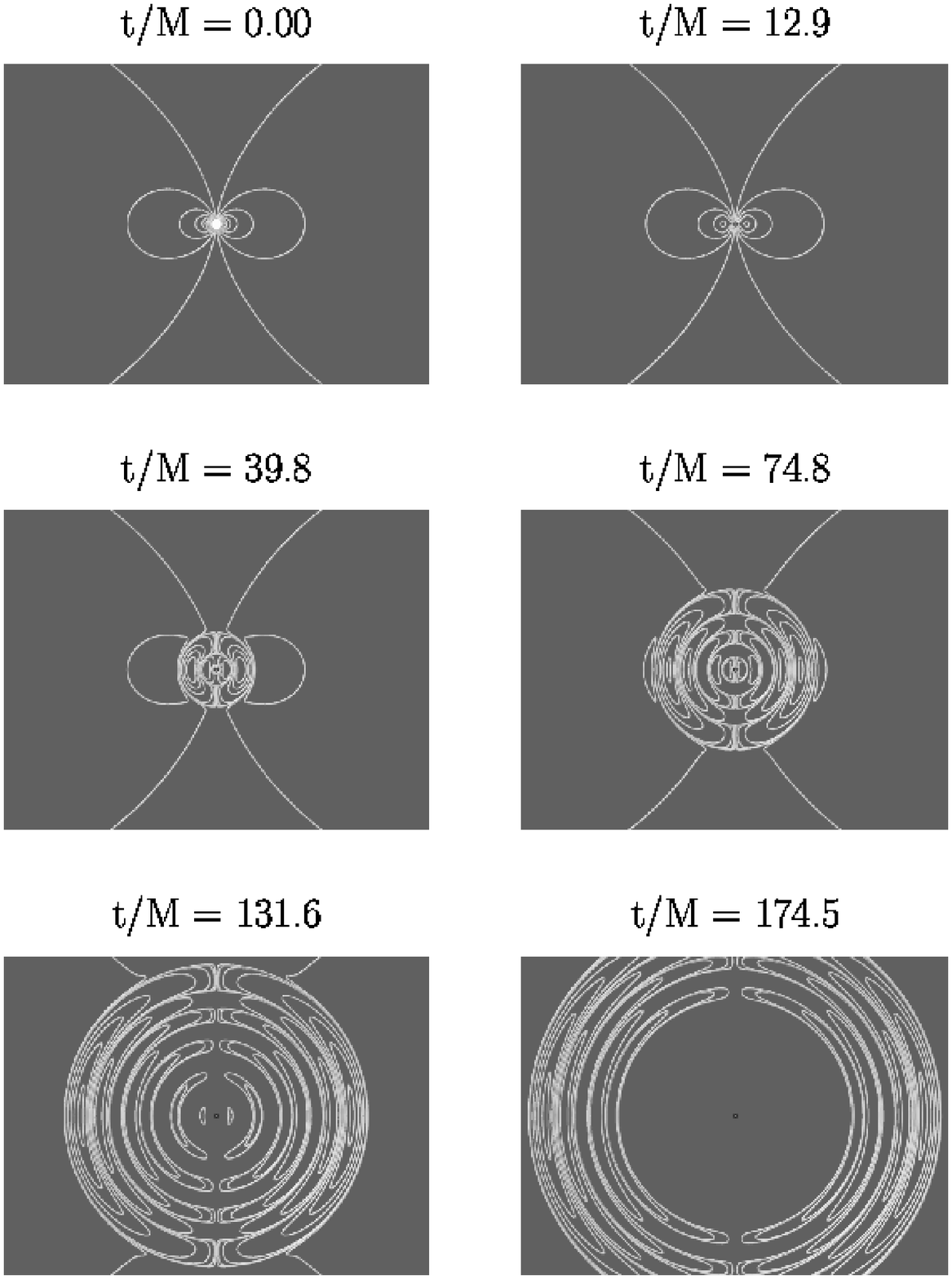}}
   \caption
     {
Snapshots of the exterior magnetic field lines at select {\it
Kerr-Schild} time slices for the same collapse depicted in Fig.
\ref{fig7}. 
The white shaded sphere covers the matter
interior; the black shaded area covers the region inside the event
horizon; the grey shaded area covers the region inside $r_s = M$
excised from the numerical grid once the matter passes inside. In this
gauge, using excision, we are able to integrate reliably to arbitrary
late times. Note
the transformation of the dipole magnetic field from a quasi-static
longitudinal to a transverse electromagnetic wave.
[From Baumgarte and Shapiro 2003b.]
  }
   \label{fig8}
   \end{figure}

This new computational tool will soon be applied to address some of the
important astrophysical problems involving GRMHD that were 
listed earlier in this section. 
To preview such applications, the 
solution to the collapse of a perfectly conducting, homogeneous, 
spherical dust ball, initially at rest and 
threaded by a weak magnetic dipole field 
has been obtained recently
(``magnetized Oppenheimer-Snyder collapse'').\cite{rf:bs03b} 
The solution applies in the limit $B^2/8\pi \ll M \rho/R$, where 
$M$ is the mass, $R$ is the radius and $\rho$ is the density of the
star. In this limit
the matter and metric are unaffected by the electromagnetic field.
Adopting this ``dynamical Cowling approximation'', the electromagnetic
field has been evolved both in the stellar interior and 
vacuum exterior. The interior solution is analytic while the exterior
requires a numerical integration. Junction conditions were used 
to match the electromagnetic fields across the stellar surface. 
This model has proven useful to experiment
with several gauge choices for handling spacetime evolution
characterized by magnetized matter, 
the formation of a black hole and the associated
appearance of a spacetime singularity. These choices ranged from ``singularity
avoiding'' time coordinates, like maximal time-slicing,
to ``horizon penetrating'' time
coordinates, like Kerr-Schild time-slicing, accompanied by black hole excision. 
The virtue of the later choice is that it
allows the integration of the exterior electromagnetic fields to proceed 
arbitrarly far into the future (see Figs~\ref{fig7} and \ref{fig8}). 
At late times the longitudinal magnetic field in the
exterior transforms into a transverse electromagnetic wave;
part of the electromagnetic radiation is captured by the hole and the
rest propagates outward to large distances. The asympotic dipole 
electromagnetic field in the exterior vacuum is found to
decay away with time as $t^{-(2l+2)}$, 
where $l = 1$, in accord with 
Price's theorem.\cite{rf:price}.  This solution 
will be used to compare with forthcoming solutions generated by 
the new GRMHD codes that will treat 
more realistic collapse scenarios. It will be particularly 
interesting to see how
the solutions will differ when the star is rotating and 
when the exterior region 
consists of a low-density, conducting atmosphere instead of a vacuum.

Simulations using the new GRMHD codes to evolve
magnetized hypermassive stars, magnetized stellar collapse, 
and the black hole - ambient disk systems they produce are now underway.
They should provide definitive answers to long-standing questions
regarding the behavior and ultimate fate of these  
systems, which seem to arise in so many astrophysical contexts.

\section*{Acknowledgements}
We would like to thank T. Baumgarte, M. Duez, J. Faber, C. Gammie, 
Y.-T. Liu, M. Shibata, B. Stephens, and K. Taniguchi 
for useful discussions. This work was
supported in part by NSF grants PHY-0205155 and PHY-0345151 and NASA
Grant NNG04GK54G at the University of Illinois.

%

\end{document}